\documentclass{article}

\usepackage{graphicx}
\usepackage{color}
\usepackage{amssymb,amsthm}

\title {How synapses can enhance sensibility of a neural network}

\begin{document}
\maketitle
\noindent{P. R. Protachevicz$^1$, F. S. Borges$^2$, K. C. Iarosz$^{2,3,*}$, I. L. Caldas$^2$, M. S. Baptista$^3$, R. L. Viana$^4$, E. L. Lameu$^5$, E. E. N. Macau$^5$, A. M. Batista$^{1,2,3,6}$}
\\
\\
\noindent {$^1$Science Post-Graduation, State University of Ponta Grossa, PR, Brazil.}\\
{$^2$Physics Institute, S\~ao Paulo University, S\~ao Paulo, SP, Brazil.}\\
{$^3$Institute for Complex Systems and Mathematical Biology, Aberdeen, UK.}\\
{$^4$Physics Department, Federal University of Paran\'a, Curitiba, PR, Brazil.}\\
{$^5$National Institute for Space Research, S\~ao Jos\'e dos Campos, SP, Brazil.}\\
{$^6$Department of Mathematics and Statistics, State University of Ponta Grossa, PR, Brazil.}

\maketitle
\begin{abstract}
\noindent In this work, we study the dynamic range in a neuronal network modelled by cellular automaton. We consider deterministic and non-deterministic rules to simulate electrical and chemical synapses. Chemical synapses have an intrinsic time-delay and are susceptible to parameter variations guided by learning Hebbian rules of behaviour. Our results show that chemical synapses can abruptly enhance sensibility of the neural network, a manifestation that can become even more predominant if learning rules of evolution are applied to the chemical synapses.
\end{abstract}




\section{Introduction}

The human brain contains about 100 billion neurons \cite{lent12} and each neuron has approximately $10^4$ connections. The connections between neurons can be mediated in terms of chemical synapses or electrical gap junctions, also known as electrical synapses \cite{bear2015}. The signal transmission in chemical synapses is unidirectional, while in electrical synapses the signal is transmitted in both directions \cite{hormuzdi04}. Furthermore, chemical synapses transmit impulses slower than electrical synapses, due to the fact that chemical synapses transfer molecules called neurotransmitters and electrical synapses transfer ionic current through the gap junction pores between neurons \cite{mamiya03}. Many different mathematical models have been used to describe dynamical behaviour of neuronal networks, such as differential equations, coupled maps, and cellular automata. Nonlinear differential equations of Hodgkin-Huxley \cite{hh52} and Hindmarsh-Rose \cite{hr84} have been considered to build neuronal networks \cite{baptista10,batista14}. The Hodgkin-Huxley neuron model is composed by four ordinary differential equations and describes the dynamics of the membrane potential by taking into account the dynamics of the ion channels. The Hindmarsh-Rose neuron is a simplified model with three ordinary differential equations that exhibits rapid firing or bursting \cite{hr}. With regard to coupled maps, there are studies about neuronal networks using the Rulkov map \cite{rulkov01} as local dynamics \cite{lameu16} in neuronal network model. It is also possible to model neuronal networks by means of cellular automata \cite{viana14}. Cellular automaton was developed by John von Neumann \cite{newmann66} and is a dynamical system with discrete time, space, and state variables \cite{wolfram84}. The study of large networks with complex evolutionary plastic rules and time-delays using limited computational power. Networks modelled by cellular automata are often considered because they are more computationally efficient than those modelled by differential equations and coupled maps.

One of the key problems in Psychophysics is the quantitative characterisation of the sensation due to a given stimulus. Stevens \cite{stevens86} proposed a stimulus-response theory, where the relationship between stimuli and response is given by a power-law. The capacity of a neuronal network to discriminate the intensity of the dynamic range, an external stimulus measured by the dynamic range. The dynamic range defined in terms of firing rates is meant to quantify the absolute ratio between the largest and smallest values of a changeable quantity as the intensity of physical stimuli, for instance, sound, light and odorant concentration high and low stimulus firing rates leading to high and low response firing rates of the neurons, respectively \cite{stevens86}. In recent work \cite{batista14}, Batista et al. found that the dynamic range increases with the network size, and therefore, the large the network the more sensitive to external stimuli it becomes \cite{batista14}.

In a network modelled with cellular automaton, Kinouchi and Copelli \cite{kinouchi06} have shown that maximisation of the analysed the optimal dynamic range electrically coupled spiking neurons could be achieved by setting neurons to interact among themselves in a critical way, where perturbations are neither damped by a group nor cascaded. Borges et al. \cite{borges16} have then investigated the dynamic range in  neural networks modelled by cellular automaton, where neurons connected both by electric and chemical synapses. They verified that the enhancement of the dynamic range depends on the proportion of electrical synapses as compare to the chemical ones. In this work, similar to the work of Ref. \cite{borges16}, we also consider a cellular automaton that describes spiking neurons in a network with connections between nearest neighbours and shortcuts corresponding to electrical synapses and chemical synapses, respectively. However, this work considers time-delays in the chemical synapses due to the fact that electrical synapses are faster than chemical synapses. Moreover, we have also included neuroplasticity in the chemical synapses to understand how learning rules of behaviour can affect the neural network sensibility.
 
Dynamical range is also a very important issue in regard to neuroplasticity. This term, also called as brain plasticity, is used to describe brain's ability to change its structure and function \cite{leone05}. The plasticity can occur due to experience, learning and memory formation \cite{kolb98}, or as a result of brain injury \cite{nakamura09}. The term plasticity was firstly introduced in neuroscience through the book entitled ``The Principles of Psychology'' written by William James in 1890 \cite{leone05}. In 1904, Santiago Ram\'on y Cajal used the term neuronal plasticity in his studies about the central nervous system \cite{stahnisch02}. Experimental evidence of plasticity was performed in 1923 by Lashley \cite{lashley23}. In 1949, Donald Olding Hebb proposed a theory about neuronal mechanisms of plasticity, known as Hebb's rule \cite{hebb49}. The rule postulates that connection between neurons is potentiated when they are actived synchronously, where the presynaptic neuron spikes before the postsynaptic neuron, while spike arrival after postsynaptic spikes leads to depression of connection.

There are many kinds of brain plasticity described in the literature. Some of them are the following: (i) presynaptic dependent scaling (PSD) \cite{mizusaki17, mizusaki12}; (ii) spike timing-dependent plasticity (STDP) \cite{borges16,borges17}; (iii) short-term plasticity (STP) \cite{zucker02}. PSD occurs on a slower time scale than STDP \cite{ibata08}. STP also has a shorter time scale and it has been studied in detail at
peripheral neuromuscular synapses \cite{purves01}.

Here, we study the effects of the Hebbian plasticity (STDP) in the dynamic range. Previous works that have investigated plasticity in the auditory system have been relevant to the determination of the dynamic range in cochlear implantation \cite{kral12,cardon12}. In this work, we show that the dynamic range in our considered network presents a hysteretic phase transition with respect to a gradual increase or decrease of the firing rate of the external perturbation, leading to an abrupt increase of the network sensibility (dynamic range) or a moderate decrease of its sensibility, respectively. This remarkable phenomenon is more likely to be found (with respect to a broader range parameters) when the network is evolved according to Hebbian rules of learning behaviour \cite{gollo12}. It was verified that bistability is related to memory maintenance \cite{fuster71}, and the path dependence to the dynamic range can be related to the time period in the olfactory system \cite{bhandawat07}.

This paper is organised as follows: In the Sec. \ref{s2} we will introduce our proposed network model of spiking neurons, and we show results of the average firing rate. Sec. \ref{s3} exhibits the dynamic range and how it is affected by plasticity. Finally, in the last Sec. \ref{s4}, we draw the conclusions.


\section{The model of spiking neurons}\label{s2}

\begin{figure}[htbp]
\begin{center}
\includegraphics[height=7cm,width=9cm]{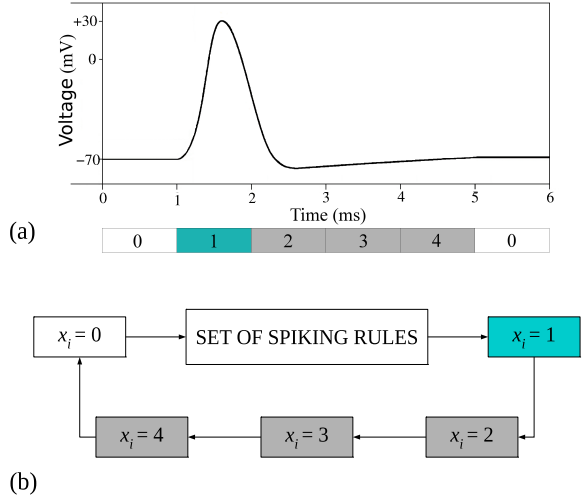}
\caption{Representation of (a) behaviour of neuron as well as the schematic diagram of the discrete state variable $x_i$ (b) cellular automaton rules.}
\label{fig1}
\end{center}
\end{figure}

A cellular automaton is built to describe a neuronal network model of spiking neurons. Figure \ref{fig1}(a) shows the shape of a typical action potential that consists of a spike upward and after a fall. We transform this behaviour into a discrete state variable $x_i$ ($x_i=0,1,2,3,4$), where $x_i=0$ is the resting state, $x_i=1$ is the spike, $x_i=2,3,4$ correspond to the refractory period. The neuron can spike when it is in the resting state $x_i=0$, however, spike does not happen during the refractory period (Fig. \ref{fig1}(b)). Excitation refers to the process of making a neuron in the resting state ($x_i=0$) to spike ($x_i=1$). Every interaction of the discrete model represents the evolution of the dynamics for a 1ms in time unit.  Therefore, a neuron changes its state in 1ms, a neuron can stay in a resting state until it is excited. Once excited, a transition for each refractory state happens in 1ms. From excitation to resting state, our model requires 4 iterations. The set of spiking rules is given by:
\begin{enumerate}
\item
A neuron $i$ can be excited by a random external stimulus that follows a Poisson distribution with average input rate $r$ \cite{borges15}. The stimulus is a detectable change in the internal or external environment \cite{craig03};
\item
A neuron $i$ with electrical synapses can be excited by an excited presynaptic neuron $j$ with probability $p_{ij}^{\rm (el)}$, where $p_{ij}^{\rm (el)}=p_{ji}^{\rm (el)}$ \cite{kinouchi06};
\item
A neuron $i$ with chemical synapses can be excited by excited neurons $j$ according to the relation $\sum_j\omega_{ij}(t-\tau)\geq T$ \cite{haimovici13}, where those presynaptic neurons that are $\omega_{ij}$ represents the connections weights between neurons $j$ to $i$, $\tau$ is the time delay, and $T$ is a threshold, i.e., the input value above which the neuron $i$ must receive to trigger.
\end{enumerate}

We construct a network with $N=10^4$ neurons, where the connections are randomly chosen \cite{erdos59}. As the control parameter of the electrical connectivity, we use the average branching ratio $\sigma=p_{ij}^{\rm(el)} K_{\rm el}$, where the probability to transmit the signal from a neuron $i$ to $j$ is given by $p_{ij}^{\rm (el)}$ $\in [0.09, 0.11]$ and $K_{\rm el}=10$ is average degree of electrical connection. For chemical connectivity, we establish a minimal threshold $T$ and average degree of chemical connection $K_{\rm ch}=5$. We also consider time delay $\tau=3$ms, chemical connections weight $\omega_{ij}=0.5$ with standard deviation $SD=0.02$.

The neuronal network response can be calculated by means of the average firing rate
\begin{equation}
F=\frac{1}{T}\sum_{t=1}^T\rho(t),
\end{equation}
where
\begin{equation} 
\rho(t)=\frac{1}{N}\sum_{i=1}^N\delta(x_i(t),1),
\end{equation}
is the density of spiking neurons, and $\delta(a,b)$ is the Kronecker delta. Figure \ref{fig2} shows the average firing rate $F$ as a function of the average input rate $r$. In Fig. \ref{fig2}(a), we see that $F$ presents a minimum ($F_{\rm min}$) and a maximum ($F_{\rm max}$) value. In addition, $F$ curve saturates for values smaller than $r_{0.1}$ and larger than $r_{0.9}$. The input rates $r_{0.1}$ and $r_{0.9}$ are obtained for $10\%$ ($F_{0.1}$) and $90\%$ ($F_{0.9}$) of the interval between $F_{\rm min}$ and $F_{\rm max}$, respectively. 

Figure \ref{fig2}(b) exhibits how the $F$ values depend on $r$ values for $T$ equal $0.50$, $0.75$, $1.00$, and $1.25$, for $\sigma=0.9$. For $T=0.75$ we verify a discontinuous phase transition and a hysteresis cycle due to bistability by varying $r_+$ upward and $r_-$ downward on the values of $F$. Gollo et al. \cite{gollo12} observed bistability in a excitable media that interact through integration of inputs received in a time interval. The interaction rule of this network correspond to the our chemical spiking rule. In our simulation, we see that bistability can also occur when chemical synapses are considered in the electrical network setting chemical integration time interval equal to 1ms. In Figs. \ref{fig2}(c) and \ref{fig2}(d) we calculate $F$ (values given by the colour code bar) for $r_+$ and $r_-$ as a function of $T$, respectively, where we consider $\sigma=0.9$. Bistability in the $F$ values is noticeable in the region $r_+\lesssim 0.14$ and $0.5\lesssim T\lesssim 1$, where the phase transition happens at about $r = 0.14$.


\section{Dynamic range}\label{s3}

\begin{figure}[htbp]
\begin{center}
\includegraphics[height=9cm,width=11cm]{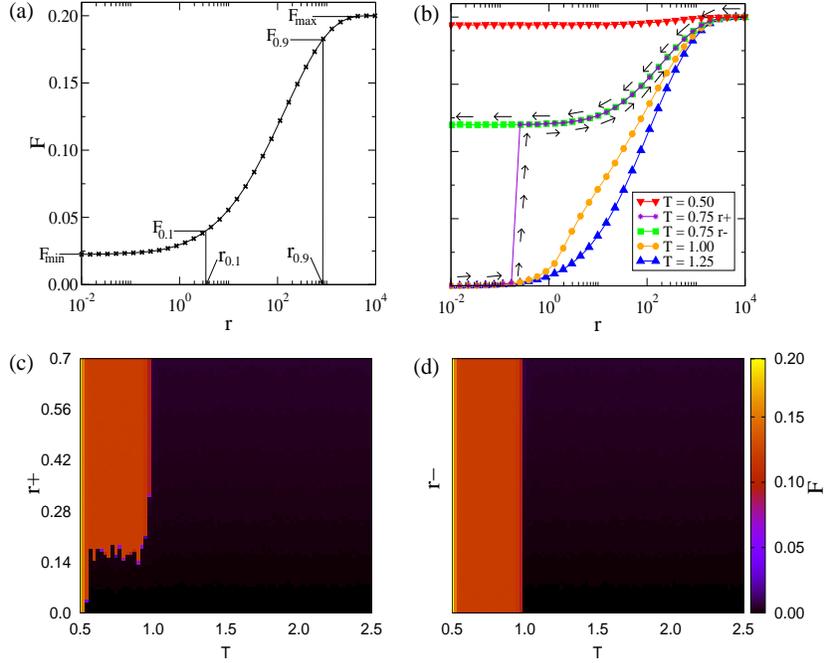}
\caption{(Colour online) The average firing rate $F$ for different configurations in the values of $T$ and $r$ in a neuronal network with $N=10^4$. In Fig. (a) , we have $\sigma=1.1$ and $T=1.5$ and we show $F$ as a function of $r$, where the input rates $r_{0.1}$ and $r_{0.9}$ are obtained for $10\%$ and $90\%$ of the interval between $F_{\rm min}$ and $F_{\rm max}$, respectively. Figure (b) exhibits results for different values of threshold $T$, while $\sigma$ is maintained constant equal to $0.9$. For $T=0.75$, $T$ values will depend on how the parameter r is varied during the emulations. Increased increments produce the values indicated by the branch $r^+$, decreasing values by the branch $r^-$. Figures (c) and (d) show the values of $F$, represented by the colour code of the side bar, for 
$\sigma=0.9$ when $r$ values are gradually incremented (c), or gradually decremented (d).}
\label{fig2}
\end{center}
\end{figure}

The dynamic range is the ratio between the largest and smallest values of a changeable quantity, and it is calculated by choosing the interval $r_{0.1}\leq r\leq r_{0.9}$, Fig. \ref{fig2}(a), in that a power-law can be fitted. It is defined as
\begin{equation}
\Delta=10\log \left(\frac{r_{0.9}}{r_{0.1}}\right),
\end{equation}
where $r_{0.1}$ and $r_{0.9}$ are obtained from $F_{0.1}$ and $F_{0.9}$, respectively. The values of $F_{0.1}$ and $F_{0.9}$ are found by means of the equation $F_x=F_{\rm min}+x(F_{\rm max}-F_{\rm min})$.

\begin{figure}[htbp]
\begin{center}
\includegraphics[height=6cm,width=8cm]{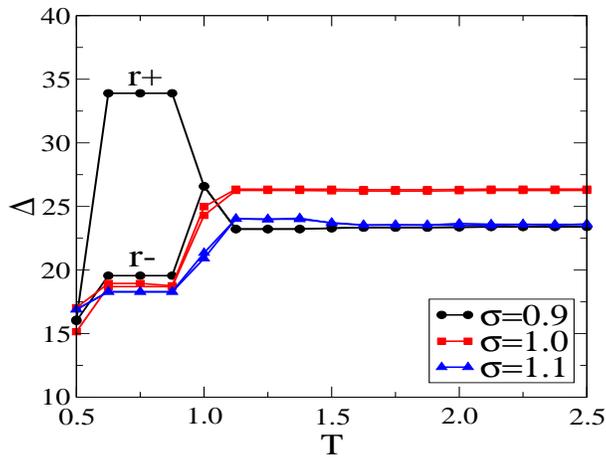}
\caption{(Colour online) Dynamic range as a function of threshold ($T$) for $\sigma=0.9$ (black circles), $\sigma=1.0$ (red squares), and $\sigma=1.1$ (blue triangles).}
\label{fig3}
\end{center}
\end{figure}

For the electrical synapses, we calculate the dynamic range for three values of $\sigma$ that correspond to values in the subcritical regime ($\sigma<1.0$), at the critical point ($\sigma=1.0$), and in the supercritical regime ($\sigma>1.0$). In the subcritical regime, the dynamic range increases with $\sigma$ until the optimal regime occurs at the critical point ($\sigma_c=1.0$). After the critical point, the dynamic range decreases and it is known as supercritical regime \cite{kinouchi06}. Figure \ref{fig3} exhibits $\Delta$ as a function of $T$ for $\sigma$ equal to $0.9$ (black circles), $1.0$ (red squares), and $1.1$ (blue triangles) for the network now presenting both electric and chemical connections. In our network, when the threshold value $T$ considered is high, the behaviour of the network is equal to network of only electrical synapses.

For $\sigma = 1.0$ the addiction of chemical synapses does not contribute to increase the value of dynamic range. For threshold values $T <1.5$, in which chemical synapses have the greatest influence in the network, the value of dynamic range decrease. For $\sigma > 1.0$, we observe a little increase of dynamic range of threshold interval $T\in[1.25,1.50]$. However, for subcritical regime, the bistable behaviour in the values of the dynamic range. This corresponds to two values of the dynamic range depending on the how the rate of the external perturbation is altered, either following the $r^+$ or the $r^-$ branches. lead to an abrupt and remarkable increase in the dynamic range external perturbation. While a gradual decrease of the average input rate ($r_-$) does not present enhancement in dynamic range, for a gradual increase ($r_+$) of external perturbation a significant increase is observed. The rate of the dynamic range following the branches maximum value about $34$dB and a minimum value about $20$dB when $r_+$ upward and $r_-$ downward, respectively.

\begin{figure}[htbp]
\begin{center}
\includegraphics[height=6cm,width=7cm]{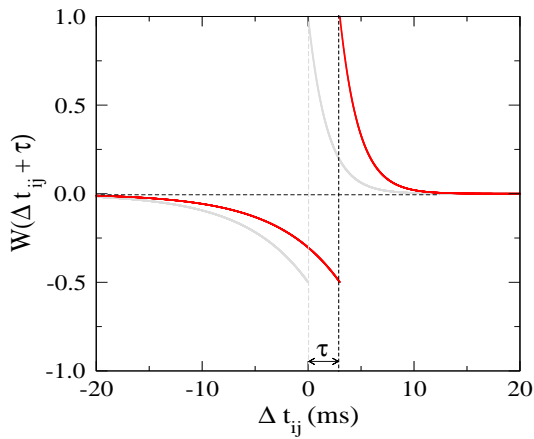}
\caption{Plasticity function (\ref{plastfunc}) as a function of the time between the spikes of the postsynaptic neuron and presynaptic, where we consider $A_1=1.0$, $A_2=0.5$, $\tau_1=1.8$ms, $\tau_2=6.0$ms and $\delta=0.001$. The gray and red lines represents $\tau=0.0$ms and  $\tau=3.0$ms, respectively.}
\label{fig4}
\end{center}
\end{figure}

With the aim at understanding the influence of the plasticity on the neuronal network modelled by cellular automaton, we consider a spike timing-dependent plasticity approach according to the Hebbian rule with time delay \cite{jalili2012}. The synaptic weights $\omega_{ij}$ are initially distributed with mean equal to $\bar \omega_{ij}=0.5$ and standard deviation equal to $SD=0.02$, they are updated in accordance with the relation
$\omega_{ij}\rightarrow \omega_{ij}+\delta W(\Delta t_{ij})$, where
\begin{equation}\label{plastfunc}
W =\left\{\begin{array}{cc}
A_1e^{-(\Delta t_{ij} -\tau) /\tau_1}, & \quad \Delta t_{ij} \geq \tau,\\
-A_2e^{(\Delta t_{ij} -\tau) /\tau_2}, & \quad \Delta t_{ij}<\tau,
\end{array} \right . 
\end{equation}
and $\Delta t_{ij}=t_i-t_j$ is the time between the spikes of the postsynaptic $t_i$ and presynaptic $t_j$ neurons. The interval of the synaptic weight is $\omega_{ij} \in [0, 1.0]$. The constant values of Eq. (\ref{plastfunc}) are $A_1 = 1.0$, $A_2 = 0.5$, $\tau_1 = 1.8$ms, $\tau_2 = 6.0$ms and $\sigma = 0.001$ \cite{popovych13}. Figure \ref{fig4} exhibits $W$ as a function of $\Delta t_{ij}$ for $\tau=0.0$ms (gray line) and $\tau=3.0$ms (red line) \cite{kandel00, northrop00}. The time delay does not change the behaviour of the curve, however, the curve is displaced according to the value of the time-delay.

\begin{figure}[htbp]
\begin{center}
\includegraphics[height=5cm,width=12cm]{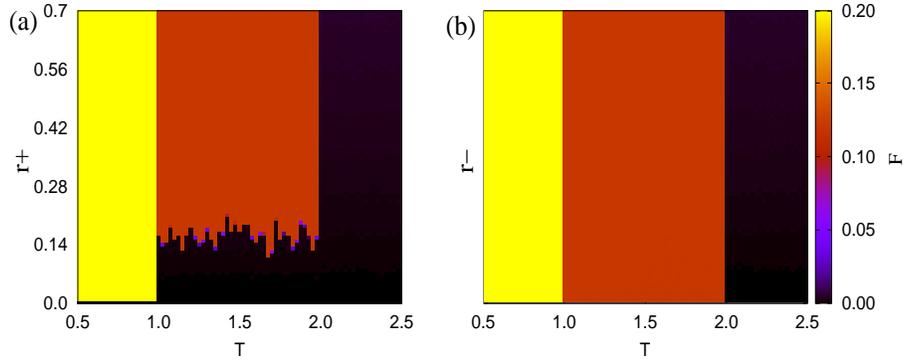}
\caption{(Colour online) Figs (a) and (b) show $r+$ and $r-$ as a function of $T$, respectively, where $\sigma=0.9$ and the colour bar represents $F$. }
\label{fig5}
\end{center}
\end{figure}

The spike timing-dependent plasticity changes the synaptic weights, and consequently changes the behaviour of the neuronal spikes. The network with STDP also presents bistability by varying $r_+$ upward and $r_-$ downward. Figure \ref{fig5}(a) and \ref{fig5}(b) shows that the bistability occurs in the region $1.0\lesssim T\lesssim 2.0$, where $\sigma=0.9$.

\begin{figure}[htbp]
\begin{center}
\includegraphics[height=6cm,width=8cm]{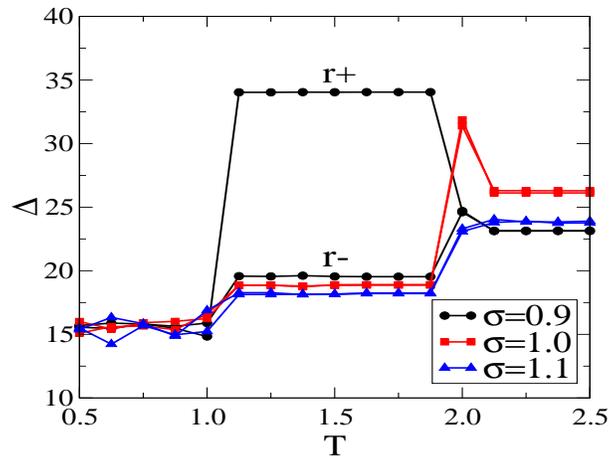}
\caption{(Colour online) $\Delta\times T$ for the network with plasticity and $T\in[0.5,2.5]$, where $\sigma=0.9$ (black circles), $\sigma=1.0$ (red squares), and $\sigma=1.1$ (blue triangles).}
\label{fig6}
\end{center}
\end{figure}

We are now ready to investigate the influence of the plasticity on the dynamic range. As a result, we verify that the network with plasticity also exhibits hysteresis, as shown in Fig. \ref{fig6}. Comparing Fig. \ref{fig6} with Fig. \ref{fig3} we observe that for $\sigma=0.9$ the maximum $\Delta$ values occur for $T>1.0$, and the region of maximum $\Delta$ is larger in the case with than without plasticity. For $\sigma=1.0$ we verified a meaningful increase of the dynamic range value only for $T=2.0$. In the case where $\sigma=1.1$, the larger value of dynamic range were found for threshold $T\ge2.125$. For these three values of $\sigma$, when $T$ is high, the behaviour of the network get close to the network behaviour with only electrical synapses. This means that learning induces sensibility to a larger range of chemical synapses, a wider $T$ value.

\section{Conclusion}\label{s4}

In this paper, we studied the dynamic range in a neuronal network modelled by a cellular automaton. We have considered networks whose neurons are connected by means of electrical and chemical synapses, the chemical connections presents time-delay, and it evolves according to Hebbian rules of learning. Our main result was to demonstrate that an enhancement of the dynamic range happens mediated by a phase transition in its value due to a hysteretic behaviour with respect to how the firing rate of the external perturbation is varied. This means that as a physical  external input perturbation becomes  larger and larger (which is encoded in the neural network by an incremental increase in the firing rate of the external perturbation), the network suddenly becomes highly sensible. On the other hand, as the external stimuli becomes  weaker and weaker, the sensibility of the neural network is only moderately reduced. This remarkable enhancement of the dynamic range happens only in the subcritical regime of the electrical network. 

 The same phase transition hysteretic behaviour was not found for the network with neurons connected only electrically in the critical or super-critical behaviour. Comparing our results with previous works \cite{viana14}, this work shows evidence that the chemical connection brings a neural network to an optimal enhanced sensibility state not achieved by the neurons if only electrically connected.

The neuronal network with plasticity also exhibits hysteresis in the subcritical regime. However, the transition occurs when the threshold $T$ value is bigger than the value for the case without plasticity. In addition, the plasticity increases the interval size of $T$ for the maximum $\Delta$. Finally, the enhancement of the dynamic range was achieved by the plasticity of the chemical synapses in the subcritical and critical regime of electrical network. Therefore, learning can enhance the  network sensibility to external perturbations.


\section*{Acknowledgements}
This work was possible by partial financial support from the following Brazilian government agencies: CNPq (154705/2016-0, 311467/2014-8), CAPES, Funda\c c\~ao Arauc\'aria, and S\~ao Paulo Research Foundation (processes FAPESP 2011/19296-1, 2015/07311-7, 2016/16148-5, 2016/23398-8, 2015/50122-0). Research supported by grant 2015/50122-0 Sao Paulo Research Foundation (FAPESP) and DFG-IRTG 1740/2



\begin{thebibliography}{00}  
\bibitem{lent12}
R. Lent, F.A.C. Azevedo, C.H. Andrade-Moraes, A.V.O. Pinto, 
How many neurons do you have? Some dogmas of quantitative neuroscience under revision, 
Eur. J. Neurosci. 35 (2012) 1-9.
\bibitem{bear2015}
M. F. Bear, B. W. Connors, M. A. Paradiso, 
Neuroscience: Exploring the Brain.
Wolters Kluwer, Philadelphia, 2015.
\bibitem{hormuzdi04}
S.G. Hormuzdia, M.A. Filippova, G. Mitropouloub, H. Monyera, R. Bruzzone,
Electrical synapses: A dynamic signaling system that shapes the activity of neuronal networks, 
Biochim. Biophys. Acta 1662 (2004) 113-137.
\bibitem{mamiya03}
A. Mamiya, Y. Manor, F. Nadim, 
Short-term dynamics of a mixed chemical and electrical synapse in a rhythmic network, 
J. Neurosci. 23 (2003) 9557-9564. 
\bibitem{hh52}
A.L. Hodgkin, A.F. Huxley, 
A quantitative description of membrane current and its application to conduction and excitation in nerve, 
J. Physiol. 117 (1952) 500-544.
\bibitem{hr84}
J.L. Hindmarsh, R.M. Rose, 
A model of neuronal bursting using three coupled first order differential equations, 
Proc. R. Soc. London B 221 (1984) 87-102.
\bibitem{baptista10}
M.S. Baptista, F.M.M. Kakmeni, C. Grebogi, 
Combined effect of chemical and electrical synapses in Hindmarsh-Rose neural networks on synchronization and
the rate of information, 
Phys. Rev. E 82 (2010) 036203.
\bibitem{batista14}
C.A.S. Batista, R.L. Viana, S.R. Lopes, A.M. Batista, 
Dynamic range in small-world networks of Hodgkin-Huxley neurons with chemical synapses, 
Physica A 410 (2014) 628-664.
\bibitem{hr}
E.J. Agnes, L.G. Brunnet, 
Model architecture for associative memory in a neural network of spiking neurons, 
Physica A, 391, (2012) 843-848.
\bibitem{rulkov01}
N.F. Rulkov, 
Regularization of synchronized chaotic bursts, 
Phys. Rev. Lett. 86 (2001) 183-186.
\bibitem{lameu16}
E.L. Lameu, F.S. Borges, R.R. Borges, A.M. Batista, M.S. Baptista, R.L. Viana,
Network and external perturbation induce burst synchronisation in cat cerebral cortex, 
Commun. Nonlinear Sci. Numer. Simulat. 34 (2016) 45-54.
\bibitem{viana14}
R.L. Viana, F.S. Borges, K.C. Iarosz, A.M. Batista, S.R. Lopes, I.L. Caldas,
Dynamic range in a neuron network with electrical and chemical synapses,
Commun. Nonlinear Sci. Numer. Simulat. 19 (2014) 164-172.
\bibitem{newmann66}
J. Von Newmann, 
Theory of self-reproducing automata, 
University of Illinois Press, London, 1966.
\bibitem{wolfram84}
S. Wolfram, 
Universality and complexity in cellular automata, 
Physica D 10 (1984) 1-35.
\bibitem{stevens86}
S.S. Stevens, L. E. Marks, 
Psychophysics: Introduction to its perceptual, neural, and social prospects, 
Transaction Publishers, Chicago, 1986.  
\bibitem{kinouchi06}
O. Kinouchi, M. Copelli, 
Optimal dynamical range of excitable networks at criticality, 
Nature Phys. 2 (2006) 348-352.  
\bibitem{borges15}
F.S. Borges, E.L. Lameu, A.M. Batista, K.C. Iarosz, M.S. Baptista, R.L. Viana,
Complementary action of chemical and electrical synapses to perception, 
Physica A 430 (2015) 236-241.
\bibitem{leone05}
A. Pascual-Leone, A. Amedi, F. Fregni, L.B. Merabet, 
The plastic human brain cortex, 
Annu. Rev. Neurosci. 28 (2005) 377-401. 
\bibitem{kolb98}
B. Kolb, I.Q. Whishaw, 
Brain plasticity and behavior, 
Annu. Rev. Psychol. 49 (1998) 43-64.
\bibitem{nakamura09}
T. Nakamura, F.G. Hillary, B.B. Biswal, 
Resting network plasticity following brain injury, 
Plos One 4 (2009) e8220.
\bibitem{stahnisch02}
F.W. Stahnisch, R. Nitsch, 
Santiago Ram\'on y Cajal's concept of neuronal plasticity: the ambiguity lives on, 
Trends Neurosci. 25 (2002) 589-591.
\bibitem{lashley23}
K.S. Lashley, 
The behavioristic interpretation of consciousness, 
Psychol. Rev. 30 (1923) 237-272.
\bibitem{hebb49}
D.O Hebb, 
The organization of behavior, 
Wiley $\&$ Sons, New York, 1949.
\bibitem{mizusaki17}
B.E.P. Mizusaki, E.J. Agnes, R. Erichsen Jr., L.G. Brunnet, 
Learning and retrieval behavior in recurrent neural networks with pre-synaptic dependent homeostatic plasticity, 
Physica A 479 (2017) 279-286.
\bibitem{mizusaki12}
B.E.P. Mizusaki, E.J. Agnes, L.G. Brunnet, R. Erichsen Jr., 
Spike timing analysis in neural networks with unsupervised synaptic plasticity, 
AIP Conf. Proc. 1510 (2012) 213-215. 
\bibitem{borges16}
R.R. Borges, F.S. Borges, E.L. Lameu, A.M. Batista, K.C. Iarosz, I.L. Caldas, R.L. Viana, M.A.F. Sanju\'an, 
Effects of the spike timing-dependent plasticity on the synchronisation in a random Hodgkin-Huxley neuronal network,
Commun. Nonlinear Sci. Numer. Simul. 34 (2016) 12-22.
\bibitem{borges17}
R.R. Borges, F.S. Borges, E.L. Lameu, A.M. Batista, K.C. Iarosz, I.L. Caldas, C.G. Antonopoulos, M.S. Baptista, 
Spike timing-dependent plasticity induces non-trivial topology in the brain, 
Neural Netw. 88 (2017) 58-64.
\bibitem{zucker02}
R.S. Zucker, W.G. Regehr, 
Short-term synaptic plasticity, 
Annu. Rev. Physiol. 64 (2002) 355-405.
\bibitem{ibata08}
K. Ibata, Q. Sun, G.G. Turrigiano, 
Rapid synaptic scaling induced by changes in postsynaptic firing, 
Neuron 57 (2008) 819-826.
\bibitem{purves01}
D. Purves, G.J. Augustine, D. Fitzpatrick, L.C. Katz, A.-S. LaMantia, J.O. McNamara, S.M. Williams, 
Neurosciende 2nd edition, Sinauer Associates, Sunderland (MA), 2001.  
\bibitem{kral12}
A. Kral, A. Sharma, 
Developmental neuroplasticity after cochlear implantation, 
Trends Neurosci. 35 (2012) 111-122.
\bibitem{cardon12}
G. Cardon, J. Campbell, A. Sharma, 
Plasticity in the developing auditory cortex: Evidence from children with sensorineural hearing loss and auditory
neuropathy spectrum disorder, 
J. Am. Acad. Audiol. 23 (2012) 396-495.
\bibitem{gollo12}
L.L. Gollo, C. Mirasso, V.M. Egu\'iluz, 
Signal integration enhances the dynamic range in neuronal systems, 
Phys. Rev. E 85 (2012) 040902.
\bibitem{fuster71}
J.M. Fuster, G.E. Alexander, Neuron activity related to short-term memory, 
Science 173 (1971) 652-654.
\bibitem{bhandawat07}
V. Bhandawat, S.R. Olsen, N.W. Gouwens, M.L. Schlief, R.I. Wilson, 
Sensory processing in the drosophila antennal lobe increases reliability and separability of ensemble odor representations, 
Nat. Neurosci. 10 (2007) 1474-1482.
\bibitem{craig03}
A. D. Craig, 
A new view of pain as a homeostatic emotion, 
Trends in Neurosci. 26 (2003) 303-307.
\bibitem{haimovici13}
A. Haimovici, E. Tagliazucchi, P. Balenzuela, D.R. Chialvo, 
Brain organization into resting state networks emerges at criticality on a model of the human connectome, 
Phys. Rev. Lett. 110 (2013) 178101.
\bibitem{erdos59}
P. Erd\"os, A. R\'enyi, 
On random graphs I, 
Publ. Math. 6 (1959) 290-297.
\bibitem{jalili2012}
M. Jalili, 
Collective behavior of interacting locally synchronized oscillations in neuronal networks, 
Commun. in Nonlinear Sci. and Num. Simul. 17 (2012) 3922-3933.
\bibitem{popovych13}
O.V. Popovych, S. Yanchuk, P.A. Tass. 
Self-organized noise resistance of oscillatory neural networks with spike timing-dependent plasticity. 
Sci. Rep. 3 (2013) 2926.
\bibitem{kandel00}
E.R. Kandel, J.H. Schwartz, T.M. Jessel, S.A. Siegelbaum, A.J. Hudspeth. 
Principles of Neural Science. 5nd edition. New York: McGraw-hill, 2012.
\bibitem{northrop00}
R.B. Northrop.
Introduction to Dynamic Modeling of Neuro-Sensory Systems. 
Biomedical Engineering. CRC Press, 2000.

\end{thebibliography}
\end{document}